\documentclass[5p,,preprint,12pt,twocolumn]{elsarticle}

\usepackage{tabulary,xcolor}
\usepackage{amsfonts,amsmath,amssymb,amsthm}
\usepackage{anyfontsize}
\usepackage[T1]{fontenc}
\usepackage[english]{babel}
\usepackage{natbib}

\makeatletter
\let\save@ps@pprintTitle\ps@pprintTitle
\def\ps@pprintTitle{\save@ps@pprintTitle\gdef\@oddfoot{\footnotesize\itshape \null\hfill\today}}
\def\hlinewd#1{%
  \noalign{\ifnum0=`}\fi\hrule \@height #1%
  \futurelet\reserved@a\@xhline}

\AtBeginDocument{\ifNAT@numbers \biboptions{sort&compress}\fi}

\makeatother

\usepackage{ifluatex}
\ifluatex
\usepackage{fontspec}
\defaultfontfeatures{Ligatures=TeX}
\usepackage[]{unicode-math}
\unimathsetup{math-style=TeX}
\else 
\usepackage[utf8]{inputenc}
\fi 
\ifluatex\else\usepackage{stmaryrd}\fi

\usepackage{url,multirow,morefloats,floatflt,cancel,tfrupee}
\makeatletter

\AtBeginDocument{\@ifpackageloaded{textcomp}{}{\usepackage{textcomp}}}
\makeatother
\usepackage{colortbl}
\usepackage{xcolor}
\usepackage{pifont}
\usepackage[nointegrals]{wasysym}
\makeatletter

\def\mcWidth#1{\csname TY@F#1\endcsname+\tabcolsep}

\def\cAlignHack{\rightskip\@flushglue\leftskip\@flushglue\parindent\z@\parfillskip\z@skip}
\def\rAlignHack{\rightskip\z@skip\leftskip\@flushglue \parindent\z@\parfillskip\z@skip}

\@ifundefined{etal}{}{}

\usepackage{ifxetex}
\ifxetex\else\if@twocolumn\@ifpackageloaded{stfloats}{}{\usepackage{dblfloatfix}}\fi\fi

\AtBeginDocument{
\expandafter\ifx\csname eqalign\endcsname\relax
\def\eqalign#1{\null\vcenter{\def\\{\cr}\openup\jot\m@th
  \ialign{\strut$\displaystyle{##}$\hfil&$\displaystyle{{}##}$\hfil
      \crcr#1\crcr}}\,}
\fi
}

\AtBeginDocument{%
  \@ifpackageloaded{endfloat}%
   {\renewcommand\efloat@iwrite[1]{\immediate\expandafter\protected@write\csname efloat@post#1\endcsname{}}}{\newif\ifefloat@tables}%
}%

\def\BreakURLText#1{\@tfor\brk@tempa:=#1\do{\brk@tempa\hskip0pt}}
\let\lt=<
\let\gt=>
\def\processVert{\ifmmode|\else\textbar\fi}

\@ifundefined{subparagraph}{
\def\subparagraph{\@startsection{paragraph}{5}{2\parindent}{0ex plus 0.1ex minus 0.1ex}%
{0ex}{\normalfont\small\itshape}}%
}{}

\newcommand\role[1]{\unskip}
\newcommand\aucollab[1]{\unskip}
  
\@ifundefined{tsGraphicsScaleX}{\gdef\tsGraphicsScaleX{1}}{}
\@ifundefined{tsGraphicsScaleY}{\gdef\tsGraphicsScaleY{.9}}{}
\def\checkGraphicsWidth{\ifdim\Gin@nat@width>\linewidth
	\tsGraphicsScaleX\linewidth\else\Gin@nat@width\fi}

\def\checkGraphicsHeight{\ifdim\Gin@nat@height>.9\textheight
	\tsGraphicsScaleY\textheight\else\Gin@nat@height\fi}

\def\fixFloatSize#1{}%\@ifundefined{processdelayedfloats}{\setbox0=\hbox{\includegraphics{#1}}\ifnum\wd0<\columnwidth\relax\renewenvironment{figure*}{\begin{figure}}{\end{figure}}\fi}{}}
\let\ts@includegraphics\includegraphics

\def\inlinegraphic[#1]#2{{\edef\@tempa{#1}\edef\baseline@shift{\ifx\@tempa\@empty0\else#1\fi}\edef\tempZ{\the\numexpr(\numexpr(\baseline@shift*\f@size/100))}\protect\raisebox{\tempZ pt}{\ts@includegraphics{#2}}}}

\AtBeginDocument{\def\includegraphics{\@ifnextchar[{\ts@includegraphics}{\ts@includegraphics[width=\checkGraphicsWidth,height=\checkGraphicsHeight,keepaspectratio]}}}

\DeclareMathAlphabet{\mathpzc}{OT1}{pzc}{m}{it}

\def\URL#1#2{\@ifundefined{href}{#2}{\href{#1}{#2}}}

\def\UrlOrds{\do\*\do\-\do\~\do\'\do\"\do\-}%
\g@addto@macro{\UrlBreaks}{\UrlOrds}

\edef\fntEncoding{\f@encoding}

\makeatother

\newif\ifmultipleabstract\multipleabstractfalse%
    \makeatletter
\def\ead{\@ifnextchar[{\@uad}{\@ead}}
\gdef\@ead#1{\bgroup
   \def\_{\string\underscorechar\space}
   \def\{{\string\lbracechar\space}
   \def\textdagger{\string\textdagger\space}
   \def\texttildeapprox{\string\texttildeapprox\space}
   \def~{\hashchar\space}
   \def\}{\string\rbracechar\space}
   \edef\tmp{\the\@eadauthor}
   \immediate\write\@auxout{\string\emailauthor
     {#1}{\expandafter\strip@prefix\meaning\tmp}}
  \egroup
}
\gdef\emailauthor#1#2{\stepcounter{ead}
      \g@addto@macro\@elseads{\raggedright
      \let\corref\@gobble
      \eadsep\texttt{#1} (#2)
      \def\eadsep{\unskip,\space}}
}

\makeatother
  
\makeatletter
\def\ps@pprintTitle{\save@ps@pprintTitle\gdef\@oddfoot{\footnotesize\hspace*{.5\textwidth}\thepage\itshape \null\hfill\today}}
\makeatother

\usepackage{booktabs}
\usepackage{float} % pour la positions des figs, tabs
\usepackage{dsfont} % pour l'indicatrice
\usepackage{comment}
\usepackage{tikz}
\usetikzlibrary{positioning, shapes, arrows}
\usetikzlibrary{shapes.geometric}
\usetikzlibrary{shapes.arrows}
\usepackage{array}
\usepackage{algorithm2e}
\usepackage{subfigure}

\DeclareMathOperator*{\argmin}{argmin}

\begin{document}

\nocite{*}

\begin{frontmatter}

    \title{
  Aging modeling and lifetime prediction of a proton exchange membrane fuel cell using an extended Kalman filter    
}
    
\author[a35d7f2a4c101,a36739788f594]{Serigne Daouda Pene\corref{c-8d81797dcb34}}
\ead{serigne-daouda.pene@laplace.univ-tlse.fr}\cortext[c-8d81797dcb34]{Corresponding author.}
\author[a35d7f2a4c101]{Antoine Picot}
%\ead{antoine.picot@laplace.univ-tlse.fr}
\author[a36739788f594]{Fabrice Gamboa}
%\ead{fabrice.gamboa@math.univ-toulouse.fr}
\author[a36739788f594]{Nicolas Savy}
%\ead{nicolas.savy@math.univ-toulouse.fr}
\author[a35d7f2a4c101]{Christophe Turpin}
%\ead{christophe.turpin@laplace.univ-tlse.fr}
\author[a35d7f2a4c101]{Amine Jaafar}
%\ead{amine.jaafar@laplace.univ-tlse.fr}
    
\address[a35d7f2a4c101]{
    Laboratoire Plasma et Conversion d'Energie, UMR 5213, CNRS, INPT, UPS \unskip, 2 rue Charles Camichel, Toulouse Cedex 7\unskip, 31071\unskip, France}
  	
\address[a36739788f594]{
    Institut de Mathématiques de Toulouse, UMR 5219, CNRS, Université Paul Sabatier\unskip, 118 route de Narbonne\unskip, Toulouse Cedex 9\unskip, 31062\unskip, France}

\begin{abstract}
This article presents a methodology that aims to model and to provide predictive capabilities for the lifetime of Proton Exchange Membrane Fuel Cell (PEMFC). The approach integrates parametric identification, dynamic modeling, and Extended Kalman Filtering (EKF). The foundation is laid with the creation of a representative aging database, emphasizing specific operating conditions. Electrochemical behavior is characterized through the identification of critical parameters. The methodology extends to capture the temporal evolution of the identified parameters. We also address challenges posed by the limiting current density through a differential analysis-based modeling technique and the detection of breakpoints. This approach, involving Monte Carlo simulations, is coupled with an EKF for predicting voltage degradation. The Remaining Useful Life (RUL) is also estimated. The results show that our approach accurately predicts future voltage and RUL with very low relative errors.
\end{abstract}

    
  \end{frontmatter}

\newif\ifdraft\drafttrue

\section{Introduction} \label{intro}
To satisfy increasingly stringent requirements in terms of $CO_2$ emissions, the Proton Exchange Membrane Fuel Cell (PEMFC) is a promising technology to decarbonize the road trafic. It will involve commercial vehicles, buses and heavy goods vehicles in the short and medium term and private cars later \cite{vadiee2015energy}. A PEMFC is an electrochemical system that directly converts chemical energy from combustion into electrical energy, heat and water. This process is based on two main equations: the oxidation of hydrogen and the reduction of oxygen. Despite the high hopes of this new technology, the inevitable aging phenomena, characterized by performance degradation over time due to changes in electrochemical and physical properties of the fuel cell components, imposes limitations on its operational lifetime and efficiency \cite{jahnke2016performance}. Performance degradation depends mainly on failures encountered during the use of the fuel cell system, losses associated with certain electrochemical phenomena and also on the operating conditions. Losses are generally categorized into four types: activation losses ($\eta_{act}$), diffusion losses ($\eta_{di\!f\!\!f}$), ohmic losses  ($\eta_{ohm}$) and parasitic losses. Activation losses reflect the kinetics of the reactions occurring in the fuel cell, with the assumption that there is no limitation due to material transport. As for diffusion losses, they result from the transport phenomena of reactants within the fuel cell. Concerning the ohmic losses, they occur in conductors and at interfaces, representing charge transport phenomena (protons and electrons). Parasitic losses encompass all phenomena that are challenging to explain and model. Issues related to the durability of fuel cell systems remain a major concern for large-scale adoption of this technology. Mechanisms for identifying aging factors, diagnosing failures and/or predicting aging must therefore be integrated into the control strategies.

The most commonly used health indicator for modeling the aging of a fuel cell is its voltage. The majority of methods attempt to establish a relationship between the voltage as output and the operating current density, as well as the operating conditions as input. Several approaches has been proposed in the literature. They are varied and grouped into three categories \cite{zhao2021review, vichard2021hybrid, hua2022review}: model-based, data-driven and hybrid methods.

%\subsection*{Model-based methods}
The objective of model-based methods is to finely describe the physical phenomena involved, or to reproduce the observed behaviour using a mathematical equation. \citet{polverino2016model} propose a semi-empirical model of the voltage of a PEMFC, primarily based on the ElectroChemical active Surface Area (ECSA), with the main objective of predicting the Remaining Useful Life (RUL). After extracting characteristic measurements of the effect of load profile on voltage degradation, \citet{zhang2017load} propose an empirical model for the lifetime prediction of an automotive PEMFC. \citet{tognan:hal-03877492} propose an empirical model based on a generic methodology dissociating the reversible and the irreversible voltage losses dynamics to predict the nominal voltage degradation with time. \citet{hu2018reconstructed} propose a voltage model considering temperature fluctuations and sensor errors for predicting PEMFC lifetime in dynamic city bus operations. \citet{zhang2012unscented} use an Unscented Kalman Filter (UKF) to track the degradations and predict the RUL of a PEMFC. \citet{liu2017prognostics} use the same semi-empirical model with an Adaptive Unscented Kalman Filter (AUKF) for RUL prediction. \citet{bressel2016remaining} introduce the Extended Kalman Filter (EKF) into the RUL prediction domain of PEMFC systems. \citet{jouin2014prognostics} explore the use of Particle Filters (PF) in the prognostic domain of PEMFC systems. The main limitation of the approaches in the literature is that the dynamical models proposed are not often linked to specific degradation phenomena. Empirical parameters are modeled with the only purpose of reproducing the voltage degradation.

%\subsection*{Data-driven methods}
Data-driven methods are generally standard statistical learning models or artificial intelligence algorithms. The dynamical behaviour of the stack is learned from appropriate measurements made on the system. They do not require knowledge of the system or physical laws, but a sufficiently large amount of data is needed to build high performance predictions. \citet{napoli2013data} employe a traditional MultiLayer Perceptron (MLP) neural network with various stacking strategies to predict the evolution of voltage and cathode temperature in a 5 kW PEMFC stack. \citet{silva2014proton} presente a methodology based on the Adaptive Neuro-Fuzzy Inference System (ANFIS) to predict temporal variations in the voltage of a PEMFC stack. The Summation-Wavelet Extreme Learning Machine (SW-ELM) method was introduced in the field of PEMFC systems by Javed et al. for the long-term prediction of RUL \cite{javed2016prognostics}. Recurrent Neural Network (RNN) architectures are currently widely used for predicting the degradation of a PEMFC system. \citet{liu2019remaining} employe a Long Short-Term Memory (LSTM) network to predict the future voltage of a proton exchange membrane fuel cell and estimate the remaining useful life. \citet{wang2020bi} propose a bidirectional LSTM incorporating an attention mechanism for predicting the degradation of the voltage in a PEMFC system. To estimate the RUL, \citet{long2022novel} use a Gated Recurrent Unit (GRU). \citet{vichard2020degradation} design an Echo State Neural Network (ESNN) model, belonging to the category of RNN, to predict the long-term performance of PEM fuel cells using 5,000 hours of experimental data.

%\subsection*{Hybrid methods}
The hybrid approach involves merging a model-based method with a data-based method. It requires a physical model describing certain degradation phenomena and a sufficient data to capture the underlying relationships not taken into account by the model. There are fewer hybrid approaches in the literature for modeling the aging of PEMFC. \citet{zhou2017degradation} integrate a Non-linear AutoRegressive Neural Network (NARNN) data-driven model with an empirical voltage model, wherein parameters are estimated using a particle filter. \citet{cheng2018hybrid} propose the Regularized Particle Filter (RPF) for three empirical models of voltage degradation combined with the Least Square Support Vector Machine (LSSVM) to predict the remaining useful life of a PEM fuel cell stack. \citet{xie2020prognostic} introduce a hybrid approach combining a particle filter and a LSTM network. With a similar approach, \citet{ma2021hybrid} combine an EKF with a LSTM. \citet{xia2022hybrid} use the locally weighted regression method to decompose the voltage data into two components: calendar aging and reversible aging. Then, an Adaptive Extended Kalman Filter (AEKF) and a long-term memory neural network are applied to predict these two parts.\newline

Our paper aims to enhance the modeling and lifetime prediction of PEMFCs by presenting a methodology that integrates parametric identification, dynamic modeling, and filtering using the extended Kalman filter algorithm. In order to develop a modeling approach, a typical aging database for a PEMFC is created under specific operating conditions. The initial step involves the characterization of the electrochemical behavior by identifying critical parameters such as exchange current ($j_0$), parasitic current ($j_n$) and limiting current ($j_{lim}$) densities. Dynamical models for the parameters of interest are proposed. These models are based on empirical observed trends. However, a significant drop is noticed in the limiting current density $j_{lim}$, leading to a change in the shape. As a result, performance declines more rapidly. This behaviour can result from a significant accumulation of water in the cell, which is one of the main technical issues in the development of PEMFC systems \cite{bhattacharya2015water}. To model such a phenomenon, we introduce an advanced modeling technique based on differential analysis to highlight critical transition points. A Monte Carlo simulation of the limiting current density is carried out to generate data according to different scenarios. The simulated curves are coupled with an extended Kalman filter on dynamical models for predicting voltage degradation. Finally, the remaining useful life is estimated by defining an End-Of-Life (EOL) criterion.

In the subsequent sections, we provide more details on our methodology. First, the simulated data are discussed. Then, the electrochemical model used is described in Section \ref{qs_model}. The parametric identification results are reported in \ref{params_id}. The proposed evolution models for the parameters of interest are described in \ref{aging_eqn}. Subsections \ref{ekf} and \ref{pred_framework} present the EKF and the framework for predicting future performance and RUL, respectively. Finally, the results are presented and discussed in Section \ref{results}.
\section{Methodology} \label{methodo}
This section outlines the steps to model the aging process and predict the lifetime. To model the aging of a fuel cell, it is crucial to have at hand a physical or mathematical equation that expresses how performance changes to specific degradation mechanisms under certain operating conditions and a given mission profile. Among various approaches, we will use a well-known model based on the polarization equation, also known as the quasi-static model. The goal is to identify the model parameters using the available data. Based on the time evolution of the parameters of interest, we propose aging laws helpful to develop a model for predicting the lifetime.

\subsection{\textbf{Data source}} \label{data_simu}

Predicting the lifetime of a fuel cell requires to have at hand data. They are used to either initialize model parameters or to learn underlying patterns. Collecting a large quantity of aging data for a fuel cell is very expensive in terms of time and resources. Numerical simulation is an alternative often deployed. To do so, we have built a typical aging database for a PEMFC under specific operating conditions. This was achieved by modeling various performance losses of a PEMFC and aging laws. The operating conditions are assured to be fixed regardless of the current over time (Table \ref{tab:COPsimu}). The targeted test duration is 38,072 hours at a fixed current density ($j = 1 A/cm^2$). In addition to these considerations, other initial assumptions are assured. More precisely: 1) there is no reversible losses in the aging process; 2) continuous and slow aging phenomena (no membrane rupture considered, for example); 3) average equivalent cell (all cells age uniformly); 4) absence of measurement noise.
\begin{table}[ht]
\footnotesize
\centering
\begin{tabular}{c c c}
\toprule
Physical parameters & Values & Units \\
\midrule
Temperature & 75 & $^\circ$C \\
Pressure & 2 & bara \\
Relative Humidity on air side & 30 & \% \\
Air side stoichiometry ($\lambda_{air}$) & 2.5 & - \\
Relative Humidity on $H_2$ Side & 50 & \% \\
$H_2$ side stoichiometry ($\lambda_{H_2}$) & 1.5 & - \\
\bottomrule
\end{tabular}
\caption{Fixed operating conditions for the simulation.}
\label{tab:COPsimu}
\end{table}

\begin{figure}[H]
    \centering
    \includegraphics[width=0.85\linewidth]{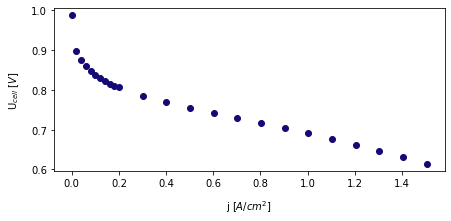}
    \caption{Initial experimental polarization curve.}
    \label{fig:initial_cpl}
\end{figure}

\begin{figure}[H]
    \centering
    \includegraphics[width=0.95\linewidth]{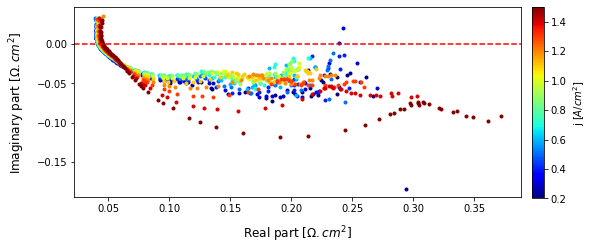}
    \caption{Experimental impedance spectroscopies for different levels of current density.}
    \label{fig:spectroscopies}
\end{figure}

To set the initial conditions for the simulation model, real-world data were employed. These data were obtained from the characterization of a stack consisting of five cells with an active surface area of 470 $cm^2$. The corresponding polarization curve and impedance spectroscopies are respectively depicted in Figure \ref{fig:initial_cpl} and \ref{fig:spectroscopies}.

\begin{figure}[ht]
    \centering
    \includegraphics[width=0.9\linewidth]{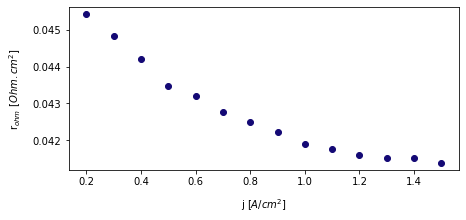}
    \caption{Evolution of the experimental ohmic resistance as a function of current density.}
    \label{fig:initial_rohm}
\end{figure}

To parameterize the polarization curve at time $t = 0$, it is important to have the values of the ohmic resistance for different levels of current density. For this, the breakpoint method with the real axis \cite{rallieres:tel-00819317} is implemented to estimate the ohmic resistance. The results obtained by applying this procedure to each impedance spectrum are depicted in Figure \ref{fig:initial_rohm}.

\begin{figure}[H]
    \centering
    \includegraphics[width=0.9\linewidth]{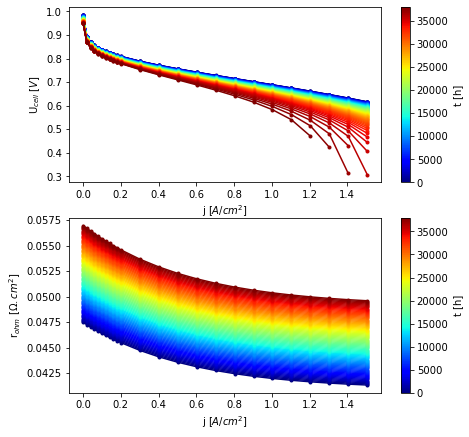}
    \caption{Simulated polarization curves (top) and ohmic resistance (bottom) as functions of current density every 500 hours.}
    \label{fig:CPL_Relec_simu}
\end{figure}

\begin{figure}[ht]
    \centering
    \includegraphics[width=\linewidth]{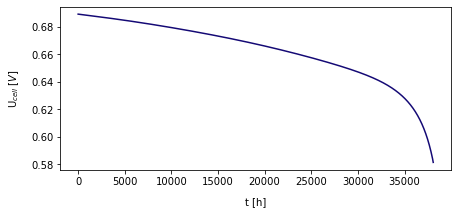}
    \caption{Simulated voltage over time, with current density fixed at $1A/cm^2$.}
    \label{fig:Ucell_t_simu}
\end{figure}

The generation of the aging database from the simulation model provides three types of information: polarization curve every 500 hours, ohmic resistance as a function of current density every 500 hours (Figure \ref{fig:CPL_Relec_simu}) and the equivalent cell voltage every hour for 38,072 hours (Figure \ref{fig:Ucell_t_simu}).

\subsection{\textbf{Electrochemical model}} \label{qs_model}
The quasi-static model describes the voltage response of a fuel cell to a current excitation. It is a powerful tool for various applications such as PEMFC characterization, aging studies, and diagnostics. This model captures losses leading to a reduction in reversible voltage ($E_{rev}$), corresponding to the theoretically obtained voltage in the absence of electrical loss. The reversible voltage $E_{rev}$, called Nernst potential \cite{mardle2021examination}, is calculated from thermodynamic quantities and depends on the temperature and pressures of $H_2$ and $O_2$.

The expression of the quasi-static model is given in Equation (\ref{eqn:eq_qsmodel}); for all $j$,
\begin{equation} \label{eqn:eq_qsmodel}
U_{cell}(j) = E_{rev} - \eta_{act}(j) - \eta_{di\!f\!\!f}(j) - \eta_{ohm}(j).
\end{equation}
This is a model with non-dissociated electrodes. However, other formulations of the quasi-static model are proposed in \cite{fontes2005, hu2018reconstructed}. This variation arises primarily from the way the losses are modeled. Here, we consider the same formulation as in \cite{aabid2020}. Hence, the losses are:
\begin{equation} \label{eqn:eta_act1}
    \eta_{act}(j) = \frac{RT}{2\alpha F} \ln \left( \frac{j+j_n}{j_0} \right),
\end{equation}
\begin{equation} \label{eqn:eta_diff1}
    \eta_{di\!f\!\!f}(j) = \frac{RT}{2\beta F} \left\lvert \ln \left( 1 - \frac{j}{j_{lim}}\right) \right \rvert,
\end{equation} 
\begin{equation} \label{eqn:eta_ohm1}
    \eta_{ohm}(j) = r_{ohm}(j) \times j. 
\end{equation}

Table \ref{tab:recap_params_qs} provides a summary of the parameters in the quasi-static model. Some of these parameters are well-known constants ($F$, $R$, $T$), while others need to be identified ($\beta$, $j_n$, $j_0$, and $j_{lim}$). The value of $\alpha$ is set at 0.5. Indeed, this assumption is commonly made when modeling a PEMFC \cite{mansouri2024investigating}.

\begin{table}[ht]
    \centering
    \resizebox{\columnwidth}{!}{%
    \begin{tabular}{c c c}
    \toprule
    Notation & Parameter & Unit\\
    \midrule
    $\alpha$ & Charge transfer coefficient & -\\
    $F$ & Faraday's constant & C/mol \\
    $R$ & Universal gas constant & J/(K.mol) \\
    $T$ & Temperature & K \\
    $j_n$ & Parasitic current density & $A/cm^2$\\
    $j_0$ & Exchange current density & $A/cm^2$\\
    $\beta$ & Diffusion coefficient & - \\
    $j_{lim}$ & Diffusion limiting current density & $A/cm^2$\\
    $r_{ohm}$ & Ohmic resistance density & $\Omega.cm^2$\\
    \bottomrule
    \end{tabular}
    }
    \caption{Parameters of the quasi-static model.}
    \label{tab:recap_params_qs}
\end{table}

\subsection{\textbf{Identification of the aging parameters}} \label{params_id}
Parametric identification is an essential step in modeling and characterizing the aging of a fuel cell. Let $\theta \in \Theta \subset \mathbb{R}^m$ be the set of the $m$ unknown parameters associated with the quasi-static model, and $N$ the number of voltage measurements at different current density. The optimal parameter $\hat{\theta}$ (wherever it exists) is given by:
\begin{equation}\label{eqn:eq_minimization}
\hat{\theta} = \argmin_{\theta \in \Theta} \sum_{i=1}^{N} \left( U_{cell}^{model, i} - U_{cell}^{data, i} \right)^2.
\end{equation}
%where $\Theta \subset \mathbb{R}^m$, here $m$ is the number of unknown parameters.

To solve this non-linear least squares problem, the Levenberg-Marquardt algorithm \cite{more2006levenberg} is used. The Root Mean Square Error (RMSE) is then calculated to assess the performance of the identification procedure.
$$RMSE = \sqrt{\frac{1}{N} \sum_{i=1}^{N} \left( \hat{U}_{cell}^{model, i} - U_{cell}^{data, i} \right)^2},$$
where $\hat{U}_{cell}^{model}$ is the estimated voltage associated with the optimal parameter $\hat{\theta}$.

\begin{figure}[H]
    \centering
    \includegraphics[width=\linewidth]{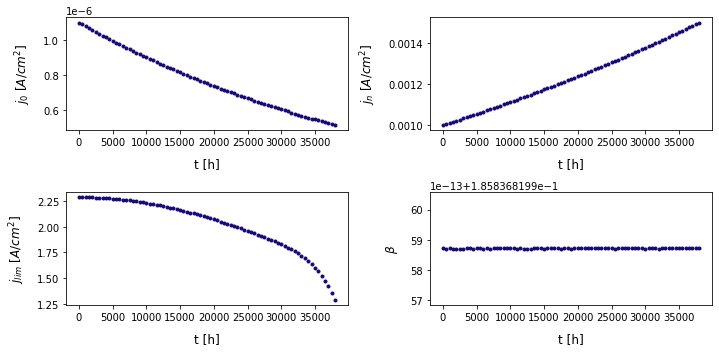}
    \caption{Evolution in time of the identified values for $j_0$, $j_n$, $j_{lim}$ and $\beta$.}
    \label{fig:model1_cas1}
\end{figure}

Initially, we consider that all the parameters $j_0$, $j_n$, $\beta$, and $j_{lim}$ vary over time. This means that they differ from one characterization to another. These four parameters are identified for each polarization curve. The evolution of the values obtained is depicted in Figure \ref{fig:model1_cas1}. The parameter $\beta$ appears to be constant over time. Thus, we assume that $\beta$ is common to all the curves. In other words, the diffusion coefficient remains constant throughout the life of the fuel cell. The trend of the other parameters remains unchanged under this assumption.

\begin{figure}[H]
    \centering
    \includegraphics[width=\linewidth]{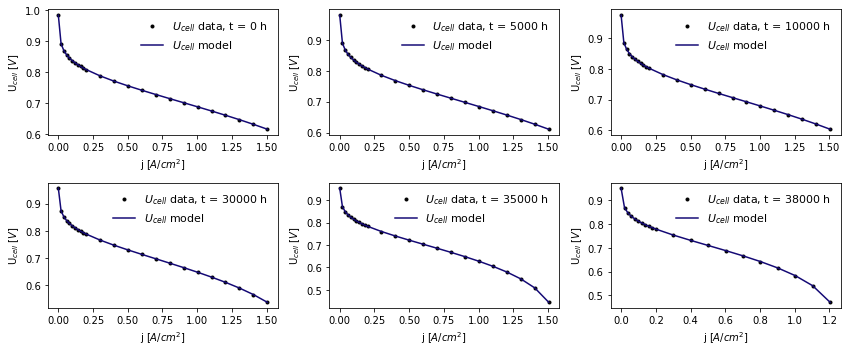}
    \caption{Simulated polarization curves and those estimated at different time points.}
    \label{fig:CPL_model1_cas2}
\end{figure}

By evaluating the voltage associated with each set of identified parameters, it appears that the obtained values lead to good adjustment of the model to the data (Figure \ref{fig:CPL_model1_cas2}).

The maximum RMSE considering all polarization curve is 9.8783$\times 10^{-17}$ which is very low.

\subsection{\textbf{Governing equations of aging}} \label{aging_eqn}

To model aging, a dynamic formulation of the quasi-static model is proposed. Indeed, the polarization curves are only produced at well-defined times (not close to each other). So, the models describing the evolution of the parameters of interest are proposed. With these models, it is then possible to assess the voltage over time.

Time-evolution models are proposed for the exchange current density $j_0$, the parasitic current density $j_n$, the limiting current density $j_{lim}$ and the ohmic resistance $r_{ohm}$.

The trends for $j_0$ and $j_n$ suggests that their evolutions can be approximated by decreasing and increasing exponential functions, respectively. So that, their expressions as a function of time are:
\begin{equation} \label{eqn:j0_t}
    j_0(t) = a_0 e^{-k_0 t},
\end{equation}
\begin{equation} \label{eqn:jn_t}
    j_n(t) = a_n e^{k_n t},
\end{equation}
where $a_0$, $a_n$, $k_0$ and $k_n$ are coefficients to estimate.

We consider a normalized version of the time defined by:
\begin{equation} \label{eqn:tnorm}
    t_{norm} = \frac{t}{t_{max}},
\end{equation}
where $t_{max}$ is set at 38,000 hours. 

Calibrating the coefficients of these models against the respective data shows that the proposed relationships effectively represent the temporal evolutions of the exchange current density and parasitic current density. The resulting mean absolute errors are 5.3627$\times 10^{-22}$ and 1.9346$\times 10^{-18}$, respectively.

\begin{figure}[H]
    \centering
    \includegraphics[width=0.9\linewidth]{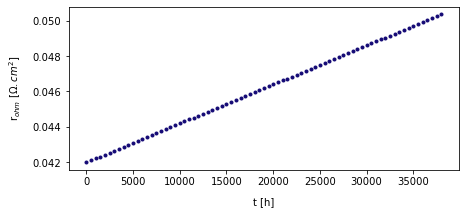}
    \caption{Temporal evolution of $r_{ohm}$ at $1 A/cm^2$.}
    \label{fig:rohm_t}
\end{figure}

Ohmic resistance values over the time are also needed to predict the future voltages. Extracting this for a current density of $1A/cm^2$ every 500 hours yields Figure \ref{fig:rohm_t}.
Obviously, a linear trend modelizes perfectly this evolution. Hence, we may write:
\begin{equation} \label{eqn:rohm_t}
r_{ohm} (t) = r_{ohm}^{0} + k_{ohm} \times t,
\end{equation}
where $r_{ohm}^{0}$ and $k_{ohm}$ are coefficients to identify.

\begin{figure}[H]
    \centering
    \includegraphics[width=0.9\linewidth]{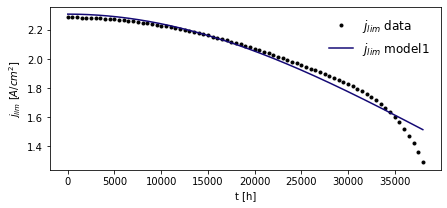}
    \caption{Fitted curves of $j_{lim}$ with model1 (Equation \ref{eqn:jlim_model1}).}
    \label{fig:jlim_model1}
\end{figure}

To model the temporal evolution of the limiting current density, we tried, in a first time, to express it as a decreasing quadratic exponential function,
\begin{equation} \label{eqn:jlim_model1}
j_{lim} (t) = a_1 e^{-k_1 t^2}.
\end{equation}
Fitting this function to the identified data gives the results depicted in Figure \ref{fig:jlim_model1}. It shows that this formulation (named model1) does not capture very well the temporal evolution of $j_{lim}$ over the entire lifetime of the system. The corresponding RMSE is 4.5831$\times 10^{-2}$.

\begin{figure}[H]
    \centering
    \includegraphics[width=\linewidth]{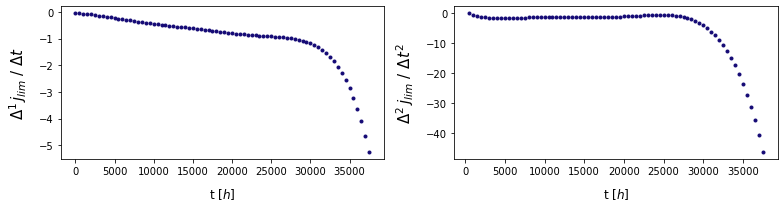}
    \caption{First and second discrete derivatives of $j_{lim}$ over time.}
    \label{fig:der_jlim}
\end{figure}

Modeling the limiting current density is more challenging. The complexity arises from the observation of a shift in the degradation regime beyond a certain point in time. This phenomenon would explain the accelerated drop in voltage after around 30,000 hours (Figure \ref{fig:Ucell_t_simu}). The change in the degradation mode of $j_{lim}$ can be elucidated by computing the first and second discrete derivatives over time (Figure \ref{fig:der_jlim}). The first discrete derivative appears to evolve linearly as a function of time in the first instants (second discrete derivative stable over the same period) and then decreases non-linearly with an increasingly negative acceleration. This clearly reflects the existence of a breakpoint in the evolution of $j_{lim}$.

\begin{figure}[H]
    \centering
    \includegraphics[width=0.9\linewidth]{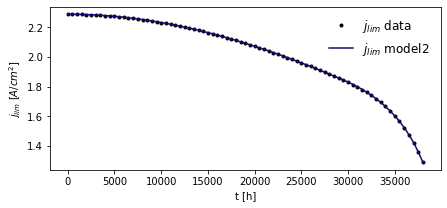}
    \caption{Fitted curves of $j_{lim}$ with model2 (Equation \ref{eqn:jlim_model2}).}
    \label{fig:jlim_model2}
\end{figure}

Taking this phenomenon into account, we propose the following model to describe the temporal evolution of $j_{lim}$:
\begin{multline} \label{eqn:jlim_model2}
    j_{lim}(t) = a_1 e^{-k_1 t^2} \\ + a_2 \left( e^{-k_2(t-t_c)^2} - 1 \right) \mathds{1}_{\{t \geq t_c\}},
\end{multline} 
where $t_c$ is the breakpoint, $a_1$, $k_1$, $a_2$, and $k_2$ are coefficients to be identified. This approach (named model2) effectively reproduce the evolution of $j_{lim}$ (Figure \ref{fig:jlim_model2}) with a RMSE of 6.5552$\times 10^{-4}$.

In the context of a predictive study, this method may have limitations. In fact, the parameters $a_2$, $k_2$, and $t_c$ of model2 may not be identifiable when one has only at hand a few data points. Indeed, accelerated degradation does not occur in the early stages of the life of the fuel cell when considering only slow and continuous aging phenomena. To overcome these drawbacks, we propose a modeling approach that accurately reflects the observed phenomena by introducing a change detection criterion in the evolution of $j_{lim}$. Before and after the change time $t_c$, the temporal evolution of $j_{lim}$ is governed by two different functions. Denoting these two functions as $P_1$ and $P_2$ respectively, we can express:
\begin{equation}\label{eqn:jlim_breakpoint}
    j_{lim} (t) = P_1(t) \times \mathds{1}_{\{t \leq t_c\}} + P_2(t) \times \mathds{1}_{\{t > t_c\}},
\end{equation}
subject to the constraint $P_1(t_c) = P_2(t_c)$ to ensure the continuity of $j_{lim}$ at the breakpoint.

Let $\tau$ be a predefined duration, and $\lambda_0$ a real number greater than 1, the change detection criterion is defined by:
\begin{equation}\label{eqn:change_detection}
\fontsize{7pt}{7pt}
t_c = \inf \left\{t-\tau :\, \left|\frac{P_1(t)-P_1(t-\tau)}{\tau}\right| \geq \lambda_0\left|\frac{P_1(t-\tau)-P_1(0)}{t-\tau}\right| \right\}.
\end{equation}

In other words, $t_c$ corresponds to the first instant the decrease in the limiting current density over the last $\tau$ hours becomes $\lambda_0$ times larger than since the beginning of the system's operation.\newline

Beyond $t_c$, the decrease in the rate of variation of $j_{lim}$ between two instants would also be more significant, providing a second derivative much more negative than what it would have been if the evolution of $j_{lim}$ had been still governed by the function $P_1$. We can express the second derivative of $P_2$ in terms of that of $P_1$ by the following relation:
\begin{equation}\label{eqn:second_derivative_link}
    P_2^{\prime\prime}(t) = \lambda P_1^{\prime\prime}(t), \,\, \forall \,\, t > t_c, \text{ with } \lambda > 1.
\end{equation}

Integrating the two sides of the equation (\ref{eqn:second_derivative_link}) twice between $t_c$ and any instant $t > t_c$, considering the continuity constraints of the function $j_{lim}$ and of its first derivative at the point $t_c$, yields the following relation:
\begin{multline} \label{eqn:P2_func}
P_2(t) = (1-\lambda) P_1(t_c) + \lambda P_1(t) \\ 
+ (1-\lambda) P_1^{\prime}(t_c)(t - t_c) \,\,\, \forall \,\, t > t_c.
\end{multline}

This modeling approach for $j_{lim}$ is used to predict the future voltage performance. More details will be given in Section \ref{pred_framework} concerning further considerations on $t_c$, the choice of $P_1$, $\lambda_0$ and $\lambda$.

\subsection{\textbf{Extended Kalman Filter}} \label{ekf}

The Kalman filter \cite{kalman1960new} is an optimal recursive filtering method widely used in various fields, including control systems, navigation, and signal processing. In statistics, filtering refers to an operation that involves estimating the state of a dynamical system from partial and noisy measurements. Such a system is described using physical or mathematical models of evolution, allowing the expression of the system's future based on past or present phenomena.

The measurements vector $Y_k$ are linked to the unobserved state $\theta_k \in \mathbb{R}^p$ ($p$ the number of state variables) by a relationship of the form:
\begin{equation}\label{obs_model}
Y_k = h_k(\theta_k, u_k) + V_k
\end{equation}
where $k \in \mathbb{N}$ is the time index, $h_k$ is the observation function, $u_k$ represents exogenous variables (external inputs) and $V_k \in \mathbb{R}^p$ is the measurement noise with covariance matrix $R_k$. This relationship is called the observation model.

The current state of the system is also expressed in terms of the past state through the state model defined by:
\begin{equation}\label{state_model}
\theta_k = f_k(\theta_{k-1}) + W_k,
\end{equation}
where $f_k$ is the transition function between two hidden states, and $W_k$ is a process noise with covariance matrix $Q_k$.

Equations (\ref{obs_model}) and (\ref{state_model}) are fundamental for filtering methods such as the Kalman filter, where the hidden state of the dynamical system is estimated by combining information from measurements and from the dynamical model. The standard Kalman filter is effective for linear systems with Gaussian noise. The extended Kalman filter \cite{ribeiro2004kalman} is an extension of the traditional Kalman filter, designed to handle non-linear dynamics and observations in a dynamical system. The non-linearities are approximated by a linearized version of the non-linear system model around the last state estimate. Then, the standard Kalman filter is applied to the linearised model.

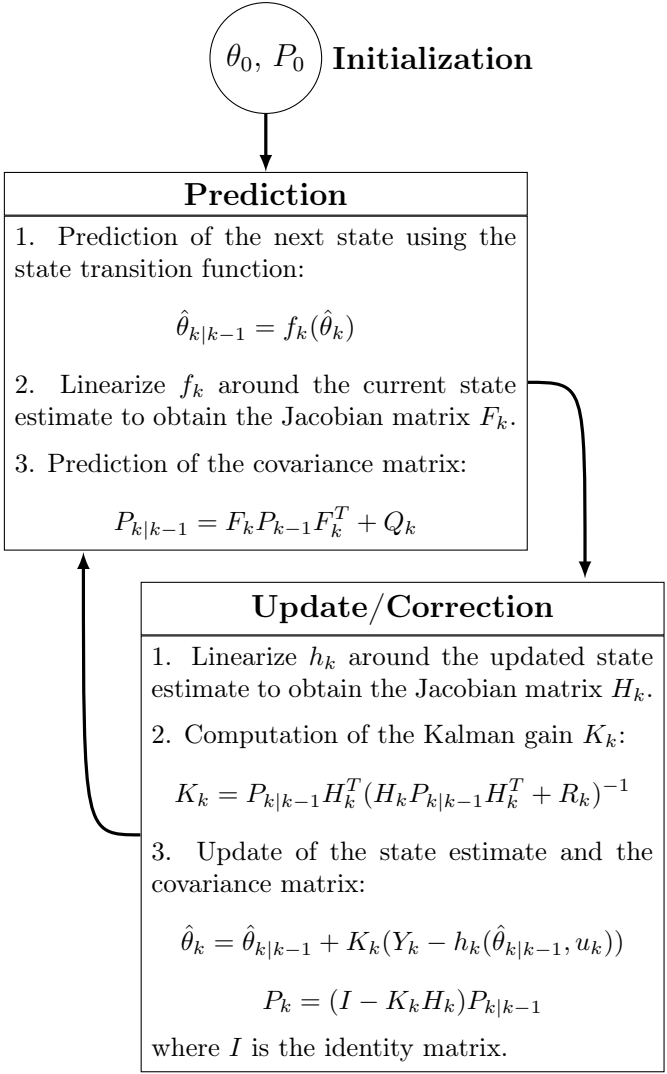
\begin{figure}[ht]
    \begin{tikzpicture}[
    node distance = 0mm,
    headernode/.style = {draw, text width=0.36\textwidth, font=\bfseries, align=center},
    textnode/.style = {draw, text width=0.36\textwidth, align=justify, font=\footnotesize},
    roundnode/.style={circle, draw}
    ]
    % initialization
    \node (textInit) [font=\bfseries] at (2.2, 1.8) {Initialization};
    \node (Init) [roundnode, align=center] at (0,1.8) {$\theta_0$, $P_0$};
    % Prediction
    \node (Pred1) [headernode] at (0,0) {Prediction};
    \node (Pred2) [draw, below=of Pred1, textnode] {
    1. Prediction of the next state using the state transition function: \[\hat{\theta}_{k|k-1} = f_k (\hat{\theta}_k)\]
    
    2. Linearize $f_k$ around the current state estimate to obtain the Jacobian matrix $F_k$.\vspace{0.2cm}
    
    3. Prediction of the covariance matrix: \[P_{k|k-1} = F_k P_{k-1} F_k^T + Q_k\]
    };
    % Correction
    \node (Corr1) [headernode] at (1.8,-5.5) {Update/Correction};
    \node (Corr2) [draw, below=of Corr1,textnode]{
    1. Linearize $h_k$ around the updated state estimate to obtain the Jacobian matrix $H_k$.\vspace{0.2cm}
    
    2. Computation of the Kalman gain $K_k$: \[K_k = P_{k|k-1} H_k^T (H_k P_{k|k-1} H_k^T + R_k)^{-1}\]
    
    3. Update of the state estimate and the covariance matrix: \[\hat{\theta}_k = \hat{\theta}_{k|k-1} + K_k (Y_k - h_k(\hat{\theta}_{k|k-1}, u_k))\] 
    \[P_k = (I-K_k H_k)P_{k|k-1}\] where $I$ is the identity matrix.
    };
    % arrows
    \tikzstyle{estun}=[->,very thick,>=latex]
    \draw[estun] (Init) to (Pred1);
    \draw[estun] (3.45,-2.5) .. controls (4.2,-2.5) and (4.2,-2.5) .. (4.2,-5.1);
    \draw[estun] (-1.65,-8.5) .. controls (-2.4,-8.5) and (-2.4,-8.5) .. (-2.4,-4.75);
\end{tikzpicture}
    \caption{Extended Kalman filter steps.}
    \label{fig:ekf_steps}
\end{figure}

The optimal values of the hidden states are computed through two main steps: prediction and update/correction (Figure \ref{fig:ekf_steps}). First, the filter must be initialized with an initial guess of the $\theta_0$ state and its covariance matrix $P_0$. The prediction and update steps are then repeated for each time step as new measurements become available. When no more measurements are available, only the prediction step is performed.

\subsection{\textbf{Lifetime prediction framework}} \label{pred_framework}

This section outlines the sequential processes developed to forecast future performance, characterized by the voltage's evolution. Subsequently, the remaining useful life will be deduced by establishing an end-of-life criterion.

Formally, the prediction of future performances can be defined as follows: given the voltage at time $t_k$, denoted as $Y_k \in \mathbb{R}$, in the sequence of measurements $Y = (Y_0, \dots, Y_k, \dots, Y_n)$ comprising $n+1$ measurements, forcast the future voltage values $\hat{Y} = (\hat{Y}_{n+1}, \dots, \hat{Y}_{n+K} )$, considering only the information available in $Y$.\newline

The proposed approach for aging prediction involves the modeling of the temporal evolution of voltage based on parameters related to different losses, namely $j_0$, $j_n$, $j_{lim}$, and $r_{ohm}$. This approach leverages both characterization data and voltage measurements over time. The concept is to use the evolution equations for the parameters $j_0$, $j_n$, $j_{lim}$, and $r_{ohm}$ as proposed in Section \ref{aging_eqn}. Subsequently, their respective coefficients are estimated using only the identified values of these parameters from polarization curves and ohmic resistance available at time $t_n$. An extrapolation is then performed to obtain the values of $j_0$, $j_n$, $j_{lim}$, and $r_{ohm}$ for all hours in the interval [0, $t_n$]. To account for uncertainties introduced by the parametric identification at two scales (polarization curves and evolution models), these extrapolations are corrected using the extended Kalman filter. Finally, the prediction of future voltage values is carried out.

Regarding the parameter $j_{lim}$, the breakpoint and how the system will degrade beyond it are unknown in principle. So that, a Monte Carlo simulation-oriented approach is adopted. The variables $t_c$ and $\lambda$ are treated as realizations of random variables. Thus, multiple scenarios for the evolution of $j_{lim}$ will be generated for $t > t_n$. For each scenario, the approach described in the previous paragraph is applied. However, two cases need to be distinguished:
\begin{itemize}
    \item Case 1: $t_c$ is not detected after implementing our time change detection algorithm applied to the values of $j_{lim}$ between 0 and $t_n$.
    \item Case 2: $t_c$ is known based on the values of $j_{lim}$ between 0 and $t_n$.
\end{itemize}

The time change detection algorithm, based on the relation (\ref{eqn:change_detection}), is described in Algorithm \ref{alg:change_detection}:

\RestyleAlgo{ruled}
\begin{algorithm} [ht]
 \caption{Time change detection} \label{alg:change_detection}
 
 \KwIn{$(j_{lim})_{t \in [t_0, tn]}$, $\tau$, $\lambda_0$}
 \KwOut{$t_c$}
 
 \tcp*[h]{\scriptsize Initialization}\;
 $t \gets 3\tau$ \;
 $\delta t \gets 1$ \;
 
 \While{$t \leq t_n$}{

 $\blacktriangleright$ Compute the absolute variation rate between $t-\tau$ and $t$\;
 \vspace{-0.2cm}
 \[\delta j_{lim}^{t:t-\tau} \gets \frac{j_{lim}(t-\tau) - j_{lim}(t)}{\tau}\]
 
 $\blacktriangleright$ Compute the absolute variation rate between $0$ and $t-\tau$\;
 \vspace{-0.2cm}
 \[\delta j_{lim}^{0:t-\tau} \gets \frac{j_{lim}(0)-j_{lim}(t-\tau)}{t-\tau}\]%
  
 $\blacktriangleright$ Compute the actual $\lambda$\;
 \vspace{-0.2cm}
 \[\lambda_{actual} \gets \frac{\delta j_{lim}^{t:t-\tau}}{\delta j_{lim}^{0:t-\tau}}\]

 \eIf{$\lambda_{actual} \geq \lambda_0$}{
   $t_c \gets t - \tau$ \;
   break \;
   }{
    $t \gets t + \delta t$\;
   }
 }
\end{algorithm}

The time window over which we inspect a change in the degradation regime is set to $\tau = 10$ hours. We also consider that there is a rupture in the limit current density when its rate of variation between $t-\tau$ and $t$ is twice that between time 0 and $t-\tau$, so $\lambda_0 = 2$. The values of $j_{lim}$ over the entire learning period ($t \in [0, t_n]$) are obtained by a cubic spline interpolation with constraints \cite{kruger2003constrained}. This ensures a relatively smooth curve with values that do not exceed intermediate values. A cubic spline associated with the family $(t_i, j_{lim}^i)$ is defined as any function $S$ belonging to class $C^2$, a polynomial of degree at most 3 in each interval $[t_i, t_{i+1}]$, and such that $S(t_i) = j_{lim}^i$ for all $i = 0, \dots, n$. Additionally, three additional conditions must be satisfied: the continuity of $S$ and of its first and second derivatives at the points $(t_i)_{0 \leq i \leq n}$.

In Case 1, we draw several values of $t_c$ from an exponential distribution. For each $t_c$, we draw a sample for $\lambda$ from a Pareto distribution. The exponential distribution is commonly used in probability theory and statistics to model a waiting time until the next event, such as success, failure, or arrival—in our case, it is the acceleration of the degradation of $j_{lim}$. The Pareto distribution is a probability distribution suitable for situations in which a small number of events or extreme values contribute significantly to the overall distribution.

Let $T_c$ be the random variable modeling the rupture instant, which follows an exponential distribution with scale parameter $\mu$. Its cumulative distribution function $F_{T_c}$ is given by:
\begin{equation} \label{eqn:fdr_exp}
F_{T_c} (t_c) = (1 - e^{-\frac{t_c}{\mu}}) \mathbb{1}_{\{t_c \geq 0\}}, \quad \mu > 0.
\end{equation}
Since $t_c > t_n$ in the first case, the values of the sample drawn from $T_c$ must be shifted by $t_n$. Furthermore, to keep only plausible values of $t_c$, we add the condition that $t_c < t_{max}$. We then have a truncated exponential distribution on $[t_n, t_{max}]$. To simulate a sample from the random variable $T_c | t_n \leq T_c \leq t_{max}$, we use the inverse Cumulative Distribution Function (CDF) method \cite{robert1999monte}. It is given by Algorithm \ref{alg:tc_simulation}.

\RestyleAlgo{ruled}
\begin{algorithm} [ht]
\caption{Simulation of a $T_c$ sample by inverse CDF method} \label{alg:tc_simulation}
\KwIn{$\mu$, $t_n$, $t_{max}$, $n_{sample}$}
\KwOut{$(t_{c,1}, \dots, t_{c,n_{sample}})$}

\For{$i\leftarrow 1$ \KwTo $n_{sample}$}{
$\blacktriangleright$ Simulate $u_i$ from a uniform distribution between 0 and 1\;
\vspace{-0.3cm}
{\[u_i \sim \mathcal{U}[0,1]\]} %\vspace{-0.7cm}
$\blacktriangleright$ Compute $t_{c,i}$ using the inverse CDF of $T_c | t_n \leq T_c \leq t_{max}$\;
\vspace{-0.5cm}
{\[t_{c,i} \gets -\mu \log\left(e^{-\frac{t_n}{\mu}} + (e^{-\frac{t_n}{\mu}} - e^{-\frac{t_{max}}{\mu}})u_i\right)\]} \vspace{-0.7cm}
}
\end{algorithm}

Let $Z$ be the random variable modeling the multiplier factor $\lambda$ of the acceleration after the rupture time. It follows a Pareto distribution with location parameter fixed to 1 and shape parameter $s$. Its CDF is given by:
\begin{equation} \label{eqn:fdr_pareto}
    F_Z (\lambda) = (1 - \lambda^{-s}) \mathbb{1}_{\{\lambda \geq 1\}}, \quad s > 0.
\end{equation}
The values of $\lambda$ are simulated using the same procedure as Algorithm \ref{alg:tc_simulation} with the relation: $$\lambda = u^{-\frac{1}{s}}, \quad u \sim \mathcal{U}[0,1].$$

The function $P_1$ (Equation (\ref{eqn:P1_func})) is defined as a polynomial of degree 2, extending the learning curve between $t_n$  and $t_c$.
\begin{equation} \label{eqn:P1_func}
    P_1 (t) = a t^2 + b t + c.
\end{equation}
To estimate the coefficients $a$, $b$ et $c$ of the function $P_1$, we solve the system of equations $S_1$:
\begin{equation}
    \footnotesize
    S_1\left \{
    \begin{array}{c @{=} c} \label{eqn:system} 
        P_1(t_n) & j_{lim}^n \\
        P_1^\prime(t_n) & S^\prime(t_n) \\
        t_c \left( P_1(t_c) - P_1(t_c+\tau) \right) & \lambda_0 \tau \left( P_1(t_c) - j_{lim}^0 \right) \\
    \end{array}
    \right.
\end{equation}
These equations enable to link $P_1$ to $S$ at the point $(t_n, j_{lim}^n)$ ensuring the continuity of the first derivative at this point. Simultaneously, it takes into account the change in the degradation mode of $j_{lim}$ through the rupture time, as specified by relation (\ref{eqn:change_detection}). The solution leads to the following expressions:
\begin{equation} \label{eqn:a_coef}
\footnotesize
    a = \frac{\lambda_0 \left( j_{lim}^0 - S^\prime(t_n)(t_c-t_n) - j_{lim}^n\right) + S^\prime(t_n)t_c}{\lambda_0(t_c-t_n)^2 - t_c(2t_c-2t_n+\tau)},
\end{equation}
\begin{equation} \label{eqn:b_coef}
    b = S^\prime(t_n) - 2 a t_n,
\end{equation}
\begin{equation} \label{eqn:c_coef}
    c = j_{lim}^n - S^\prime(t_n) t_n + a t_n^2.
\end{equation}
After $t_c$, the function $P_2$ (Equation (\ref{eqn:P2_func})) is used with the simulated values of $t_c$ and $\lambda$ to continue the prediction.

In the second case, that is, when $t_c$ is detected through Algorithm \ref{alg:change_detection}, we have $t_c < t_n$. Only a sample of $\lambda$ will be used to simulate various scenarios anticipating how the acceleration of degradation will appear. The estimated cubic spline $S$ over the interval $[0, t_n]$ is used replacing the function $P_1$ proposed in case 1. To predict the values of $j_{lim}$ beyond $t_n$, the expression of the function $P_2$ is used with $t_n$ instead of $t_c$ and $S$ instead of $P_1$.

\begin{figure}[H]
    \centering
    \subfigure{\includegraphics[width=0.48\linewidth]{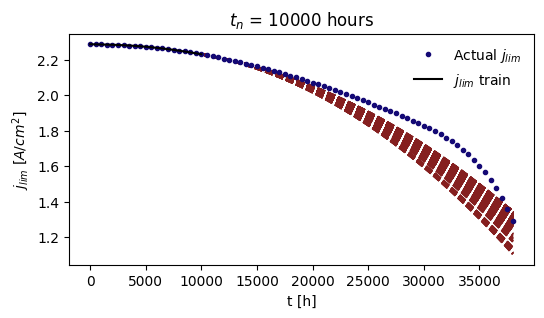}} 
    \subfigure{\includegraphics[width=0.48\linewidth]{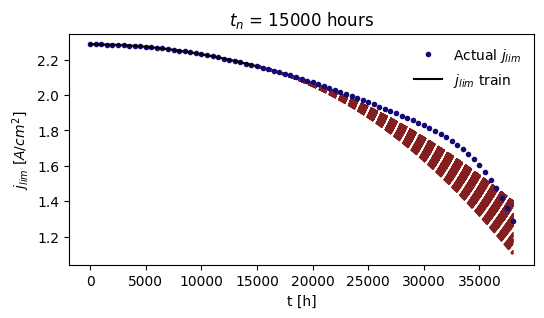}} 
    \subfigure{\includegraphics[width=0.49\linewidth]{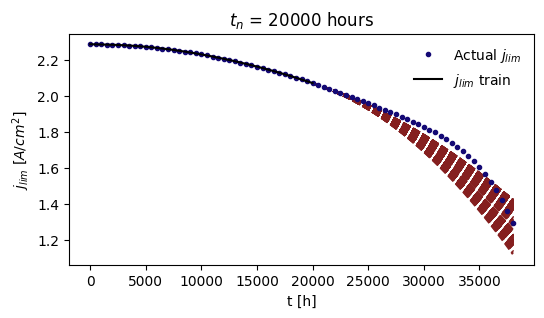}}
    \subfigure{\includegraphics[width=0.49\linewidth]{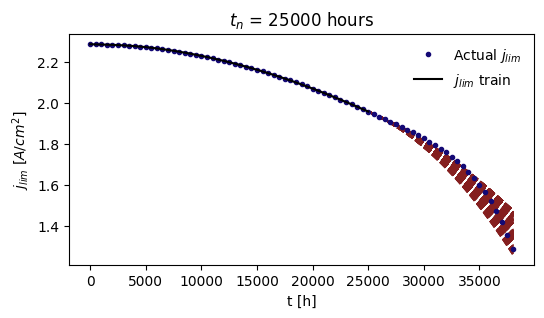}}
    \subfigure{\includegraphics[width=0.49\linewidth]{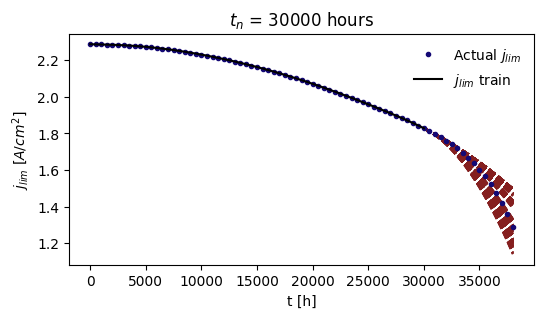}}
    \subfigure{\includegraphics[width=0.49\linewidth]{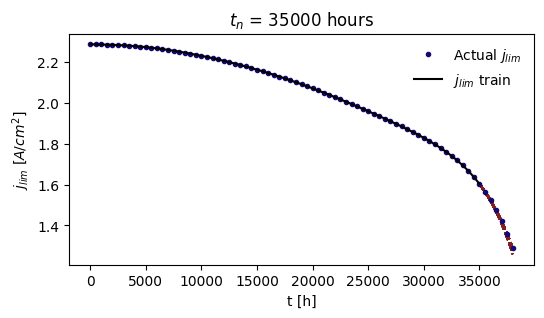}}
    \caption{Simulation of 500 $j_{lim}$ curves for different learning durations. The red curves represent the simuled scenarios.}
    \label{fig:jlim_simu_examples}
\end{figure}

The figure \ref{fig:jlim_simu_examples} depicts some examples of $j_{lim}$ simulations for different learning durations.

Given this modeling approach for $j_{lim}$, this parameter is considered as an exogenous variable in the EKF. For each simulated $j_{lim}$ curve, the EKF is implemented for voltage prediction.

Let $\theta_k = \begin{bmatrix} j_0^k & j_n^k & r_{ohm}^k \end{bmatrix}^T \in \mathbb{R}^3$ be the state vector. The state model is given by: 
\begin{equation} \label{eqn:model_state}
    \theta_{k+1} = \mathbf{F}_k \theta_k + \mathbf{f}_k + W_k,
\end{equation}
with $\mathbf{F}_k = \hspace{-0.15cm}\begin{bmatrix} e^{-k_0 \Delta t} & 0 & 0\\ 0 & e^{k_n \Delta t} & 0 \\ 0 & 0 & 1 \end{bmatrix}$ and $\mathbf{f}_k = \footnotesize \begin{bmatrix} 0\\ 0\\ k_{ohm}\Delta t \end{bmatrix}.$

As for the observation model, it is given by:
\begin{equation} \label{eqn:model_observation}
    Y_k  = h(\theta_k, j_{lim}^k) + V_k,
\end{equation}
\vspace{-0.4cm}
where {\footnotesize \begin{multline*}
     h(\theta_k, j_{lim}^k) = E_{rev} - \frac{RT}{F} \ln \left( \frac{j+j_n^k}{j_0^k} \right) - r_{ohm}^k(j) \times j \\+ \frac{RT}{2\beta F} \ln \left( 1 - \frac{j}{j_{lim}^k}\right).
\end{multline*}}

As the relationship between the measurements and the state variables is nonlinear, the Jacobian matrix of the observation function must be computed. It is given by:
$$
\begin{aligned}
    H_k &= \frac{\partial h(\theta_k , j_{lim}^k)}{\partial \theta_k} , \\
        &= \large \begin{bmatrix} \frac{RT}{Fj_0^k} & - \frac{RT}{F(j + j_n^k)} & j \end{bmatrix}.
\end{aligned}
$$

The current density $j$ is set to 1 $A/cm^2$, corresponding to the current profile considered to simulate the aging.

Once the future voltage values are predicted, the RUL relative to the learning period will be estimated. To achieve this, the end-of-life criterion considered is the loss of 10 \% in performance at 1 $A/cm^2$. The end-of-life instant, denoted $t_{EOL}$, is therefore defined as the first instant at which the threshold is reached.
\begin{equation} \label{eqn:t_fail}
    t_{EOL} = \inf\{t_k, k \in \mathbb{N} :\, Y_k \leq 0.90 \times Y_0\}.
\end{equation}

The formula for RUL in respect of $t_n$ is given by:
\begin{equation} \label{eqn:rul}
    \operatorname{RUL} = t_{EOL} - t_n.
\end{equation}
In order to evaluate the performance of the approach the Absolute Percentage Error (APE)
\begin{equation} \label{eqn:ape}
    \operatorname{APE} = 100 \times \left\lvert \frac{\operatorname{RUL}_{actual} - \operatorname{RUL}_{estimate}}{\operatorname{RUL}_{actual}}\right\rvert,
\end{equation}
will be calculated.

\section{Results and discussion} \label{results}

In this section, the estimates of the state variables by EKF for several learning durations are discussed. The predictions of the voltage for different scenarios of the evolution of $j_{lim}$ are also examined. We end with the estimates of the corresponding remaining useful life. The results obtained by implementing our approach are compared with those given by the application of an EKF considering Equation (\ref{eqn:jlim_model1}) (model1 for $j_{lim}$).

\subsection{State variables and voltage prediction} \label{state_var_voltage_prediction}

\begin{figure}[H]
    \centering
    \includegraphics[width=\linewidth]{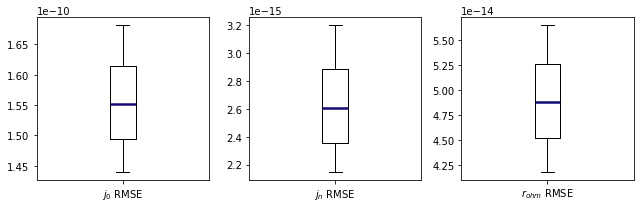}
    \caption{Distribution of the root mean square error of state variable estimates.}
    \label{fig:rmse_state_var}
\end{figure}

The state variables $j_0$, $j_n$ and $r_{ohm}$ are well predicted by the EKF algorithm whatever the learning time $t_n$ with a low error. Figure \ref{fig:rmse_state_var} illustrates the distribution of the RMSE of the estimates of the state variables. On average, the RMSE for parameter $j_0$ is 1.5549$\times 10^{-10}$, 2.6311$\times 10^{-15}$ for $j_n$ and 4.8949$\times 10^{-14}$ for $r_{ohm}$.

\begin{figure}[H]
    \centering
    \subfigure{\includegraphics[width=0.48\linewidth]{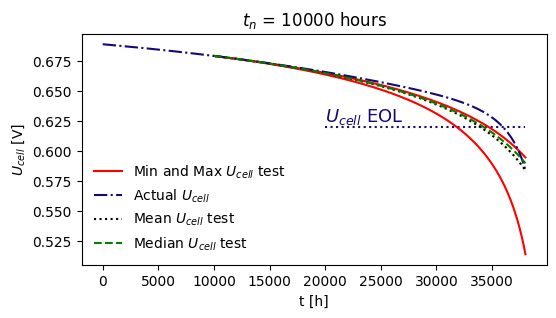}} 
    \subfigure{\includegraphics[width=0.48\linewidth]{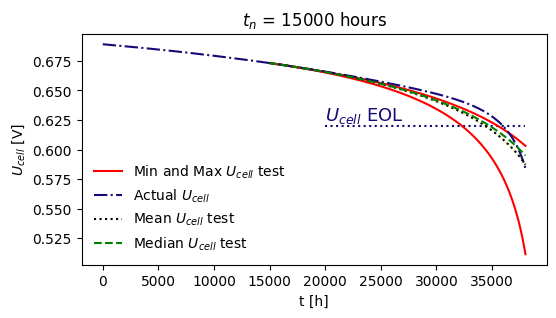}} 
    \subfigure{\includegraphics[width=0.49\linewidth]{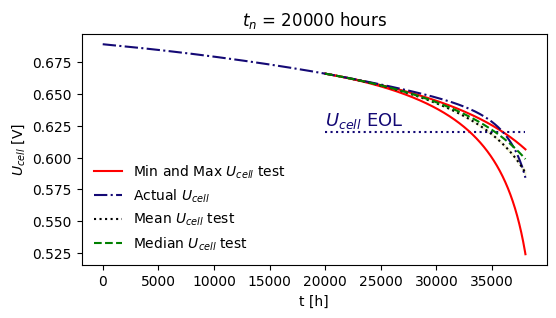}}
    \subfigure{\includegraphics[width=0.49\linewidth]{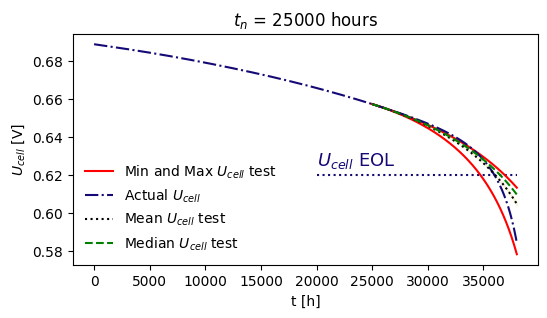}}
    \subfigure{\includegraphics[width=0.49\linewidth]{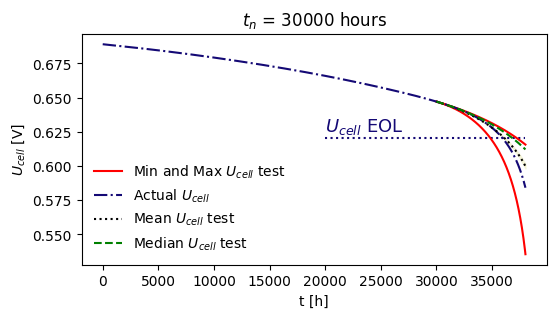}}
    \subfigure{\includegraphics[width=0.49\linewidth]{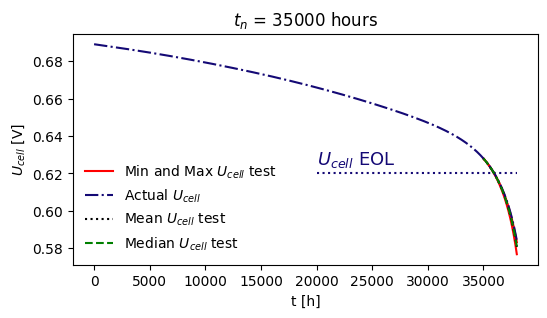}}
    \caption{Statistics on the distribution of the voltage prediction by EKF for different learning duration.}
    \label{fig:Upred_mean_median_ekf_examples}
\end{figure}

\begin{figure}[H]
    \centering
    \includegraphics[width=\linewidth]{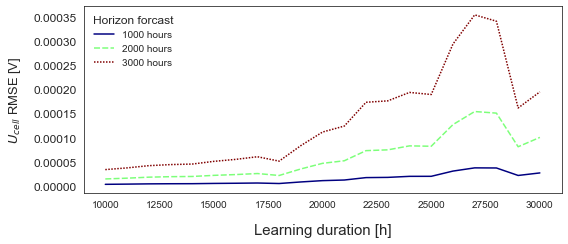}
    \caption{Root mean square error at different forecast horizons for different learning durations.}
    \label{fig:rmse_horizon}
\end{figure}

Figure \ref{fig:Upred_mean_median_ekf_examples} gives some statistics on the distribution of voltage predictions for 500 simulations of $j_{lim}$ at different learning durations. It can be seen that the proposed methodology provides good voltage predictions at the nearest horizons, whatever the learning times are. For example, the root mean square errors calculated over different horizons of the first 3,000 future hours are relatively low, with a maximum value of 3.5438$\times 10^{-4}$ (Figure \ref{fig:rmse_horizon}).

The increase in errors as a function of learning duration can be explained by the fact that the behaviour of the voltage is more difficult to predict at higher instants due to the regime change in the limiting current density $j_{lim}$. But, the longer the training period, the closer we get to the shape of the voltage at higher instants. After a certain learning duration, the predicted voltage curves form an envelope for the true values.

\begin{figure}[H]
    \centering
    \includegraphics[width=\linewidth, clip=true,trim = 0 0 0 87]{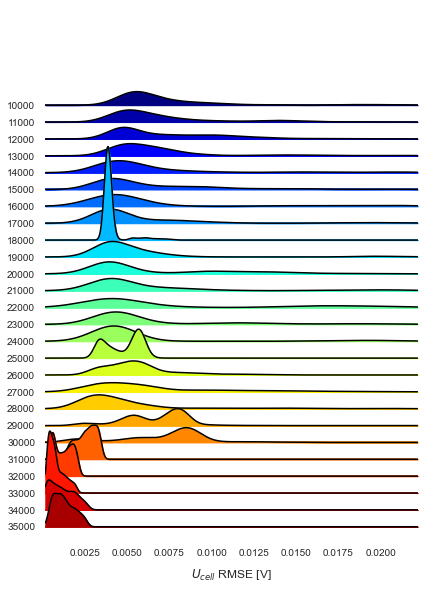}
    \caption{Distribution of the root mean square error of the voltage for different learning durations.}
    \label{fig:rmse_upred}
\end{figure}

The prediction errors decrease overall with increasing learning time and are relatively low (Figure \ref{fig:rmse_upred}). The distribution of RMSE shifts progressively towards the lowest values. The minimum and maximum mean RMSE are respectively 7.9997$\times 10^{-4}$ and 7.5273$\times 10^{-3}$. For training durations ranging from 20,000 to 30,000 hours, there is a greater dispersion of errors, in the form of a mixture of two distributions. The envelope created by all the simulations would therefore not be distributed homogeneously around the true voltage values for these $t_n$ values.

\begin{figure}[H]
    \centering
    \includegraphics[width=0.9\linewidth]{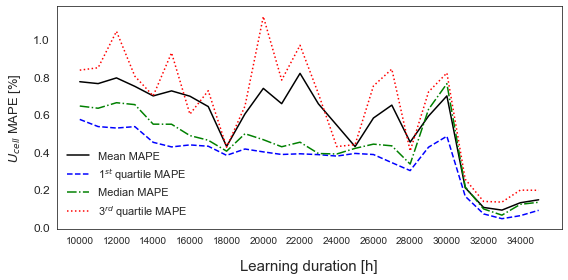}
    \caption{Statistics on the distribution of the mean absolute percentage error (MAPE) of the voltage for different learning durations.}
    \label{fig:stat_mape_upred}
\end{figure}

In Figure \ref{fig:stat_mape_upred}, the trends of some statistics on the distribution of mean relative absolute errors are illustrated. On average, MAPE varies from 0.8179\% to 0.4290\% for learning times between 10,000 and 30,000 hours. Beyond that, the average MAPE is lower, ranging from 0.2092\% to 0.0904\%. In fact, this range contains the instants from which the failure time $t_c$ is detected. Thus, this makes possible to better predict the behaviour of the curve. Its value is equal to 30,181 hours. The evolution of the median MAPE indicates that 50\% of the predicted curves gives MAPEs between 0.6617\% and 0.0635\% depending on the learning duration except for $t_n = $ 30,000 hours where the median MAPE is 0.7629\%.

\subsection{RUL estimation}\label{rul_estimation}
Remaining useful life is a widely used prognostic indicator for PEMFC systems. As defined in Section \ref{pred_framework} Equation (\ref{eqn:t_fail}), the EOL time is first estimated for all predicted curves at different training durations, then the RUL is deduced. The true value of the EOL time is 35,966 hours according to the criterion considered.

\begin{figure}[H]
    \centering
    \includegraphics[width=\linewidth]{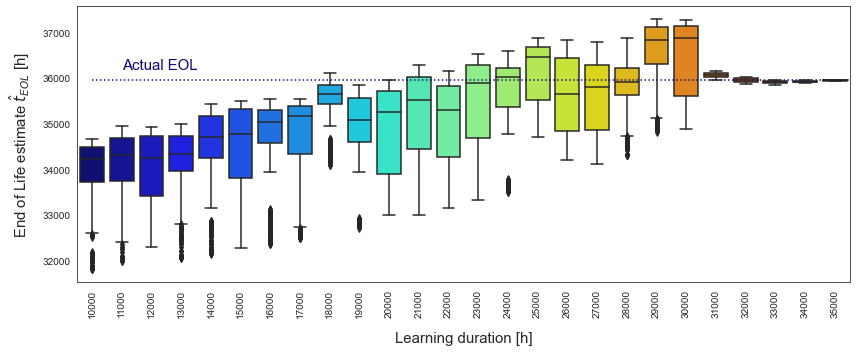}
    \caption{Distribution of the end-of-life estimate for different learning durations.}
    \label{fig:tEOL}
\end{figure}

Figure \ref{fig:tEOL} shows that as the learning duration increases, the estimates of the end-of-life time are closer to the true value. For $t_n$ between 23,000 and 30,000 hours, some predictions tend to overestimate the end-of-life time. Despite this, the estimated RUL remains closer to the true value.

\begin{figure}[ht]
    \centering
    \includegraphics[width=0.9\linewidth]{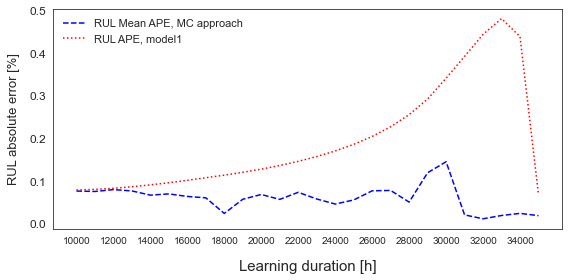}
    \caption{Comparison of RUL absolute percentage error for two methods at different learning durations.}
    \label{fig:RUL_APE}
\end{figure}

On average, the absolute error of the predicted RUL varies from 2,084 hours to 20 hours between 10,000 and 35,000 training hours. Given the training time and the prediction horizons, the absolute errors obtained on the RUL estimate are relatively small. Figure \ref{fig:RUL_APE} shows the mean absolute percentage error of the RUL predicted with the proposed approach compared to the APE obtained by applying an EKF considering Equation (\ref{eqn:jlim_model1}) (model1) for the evolution of $j_{lim}$. It is clear that the proposed approach is more effective in predicting the remaining useful life. The maximum mean APE of our approach is 0.2185\% for all learning times included, which is three times less than that obtained with model1. 

\section{Conclusions and future works} \label{conclusion}

This paper has proposed an innovative methodology that integrates parametric identification, dynamical modeling, and filtering using the Extended Kalman Filter algorithm for aging modeling and lifetime prediction of a PEMFC. Given that PEMFC are commonly utilized in complex systems, it is crucial to monitor and predict their health state with precision to facilitate the implementation of necessary measures for maintaining system integrity and reliability. Particular attention was done on modeling the limiting current density by introducing an advanced model based on differential analysis to highlight critical transition points. For voltage degradation prediction, a Monte Carlo simulation of the limiting current density was conducted. Results show that the proposed method can accurately predict the voltage with a maximum mean MAPE of 0.8179\%. The estimation of the RUL generates a maximum APE of 0.2185\%. However, all this is based on a simulated database in the absence of measurement noise. For the perspectives, we will apply the proposed approach to a real aging database representative of real-world automotive operating conditions.

\section*{Acknowledgements}
This work is part of the ECH2 project, financed by the French Government as part of the France 2030 plan operated by ADEME. The consortium is composed of: Helion ALSTOM Hydrogen SAS, Siemens Digital Industries Software, IFP Energies Nouvelles, Vitesco Techonologies, Laboratoire Plasma et Conversion d'Energie and Institut de Mathématiques de Toulouse.

\bibliographystyle{unsrtnat}

\bibliography{references}

\begin{thebibliography}{37}
\providecommand{\natexlab}[1]{#1}
\providecommand{\url}[1]{\texttt{#1}}
\expandafter\ifx\csname urlstyle\endcsname\relax
  \providecommand{\doi}[1]{doi: #1}\else
  \providecommand{\doi}{doi: \begingroup \urlstyle{rm}\Url}\fi

\bibitem[Vadiee et~al.(2015)Vadiee, Yaghoubi, Sardella, and
  Farjam]{vadiee2015energy}
Amir Vadiee, Mahmoud Yaghoubi, Marco Sardella, and Pardis Farjam.
\newblock Energy analysis of fuel cell system for commercial greenhouse
  application--a feasibility study.
\newblock \emph{Energy Conversion and Management}, 89:\penalty0 925--932, 2015.

\bibitem[Jahnke et~al.(2016)Jahnke, Futter, Latz, Malkow, Papakonstantinou,
  Tsotridis, Schott, G{\'e}rard, Quinaud, Quiroga,
  et~al.]{jahnke2016performance}
Thomas Jahnke, Georg Futter, Arnulf Latz, Thomas Malkow, Georgios
  Papakonstantinou, Georgios Tsotridis, Pascal Schott, Mathias G{\'e}rard,
  Manuelle Quinaud, Matias Quiroga, et~al.
\newblock Performance and degradation of proton exchange membrane fuel cells:
  State of the art in modeling from atomistic to system scale.
\newblock \emph{Journal of Power Sources}, 304:\penalty0 207--233, 2016.

\bibitem[Zhao et~al.(2021)Zhao, Li, Shum, and McPhee]{zhao2021review}
Jian Zhao, Xianguo Li, Chris Shum, and John McPhee.
\newblock A review of physics-based and data-driven models for real-time
  control of polymer electrolyte membrane fuel cells.
\newblock \emph{Energy and AI}, 6:\penalty0 100114, 2021.

\bibitem[Vichard et~al.(2021)Vichard, Steiner, Zerhouni, and
  Hissel]{vichard2021hybrid}
Lo{\"\i}c Vichard, N~Yousfi Steiner, Noureddine Zerhouni, and Daniel Hissel.
\newblock Hybrid fuel cell system degradation modeling methods: A comprehensive
  review.
\newblock \emph{Journal of Power Sources}, 506:\penalty0 230071, 2021.

\bibitem[Hua et~al.(2022)Hua, Zheng, Pahon, P{\'e}ra, and Gao]{hua2022review}
Zhiguang Hua, Zhixue Zheng, Elodie Pahon, Marie-C{\'e}cile P{\'e}ra, and Fei
  Gao.
\newblock A review on lifetime prediction of proton exchange membrane fuel
  cells system.
\newblock \emph{Journal of Power Sources}, 529:\penalty0 231256, 2022.

\bibitem[Polverino and Pianese(2016)]{polverino2016model}
Pierpaolo Polverino and Cesare Pianese.
\newblock Model-based prognostic algorithm for online rul estimation of pemfcs.
\newblock In \emph{2016 3rd Conference on Control and Fault-Tolerant Systems
  (SysTol)}, pages 599--604. IEEE, 2016.

\bibitem[Zhang et~al.(2017)Zhang, Yang, Luo, and Dong]{zhang2017load}
Xinfeng Zhang, Daijun Yang, Minghui Luo, and Zuomin Dong.
\newblock Load profile based empirical model for the lifetime prediction of an
  automotive pem fuel cell.
\newblock \emph{International Journal of Hydrogen Energy}, 42\penalty0
  (16):\penalty0 11868--11878, 2017.

\bibitem[Tognan et~al.(2017)Tognan, Turpin, Ralli{\`e}res, Verdu, Lombard, and
  Rakotondrainibe]{tognan:hal-03877492}
Malik Tognan, Christophe Turpin, Olivier Ralli{\`e}res, O.~Verdu, K.~Lombard,
  and A.~Rakotondrainibe.
\newblock {Analysis and Modeling of the performance degradation dynamics of a
  H2/O2 PEM Fuel Cell stack under constant current solicitation}.
\newblock In \emph{{The 12th International Conference on Modeling and
  Simulation of Electric Machines, Converters and Systems (ELECTRIMACS'2017),
  July 4-6, 2017, Toulouse (FRANCE)}}, Toulouse, France, 2017.
\newblock URL \url{https://hal.science/hal-03877492}.

\bibitem[Hu et~al.(2018)Hu, Xu, Li, Ouyang, Song, and
  Huang]{hu2018reconstructed}
Zunyan Hu, Liangfei Xu, Jianqiu Li, Minggao Ouyang, Ziyou Song, and Haiyan
  Huang.
\newblock A reconstructed fuel cell life-prediction model for a fuel cell
  hybrid city bus.
\newblock \emph{Energy conversion and management}, 156:\penalty0 723--732,
  2018.

\bibitem[Zhang and Pisu(2012)]{zhang2012unscented}
Xian Zhang and Pierluigi Pisu.
\newblock An unscented kalman filter based approach for the healthmonitoring
  and prognostics of a polymer electrolyte membrane fuel cel.
\newblock In \emph{Annual Conference of the PHM Society}, volume~4, 2012.

\bibitem[Liu et~al.(2017)Liu, Chen, Zhu, Su, and Hou]{liu2017prognostics}
Hao Liu, Jian Chen, Chuyan Zhu, Hongye Su, and Ming Hou.
\newblock Prognostics of proton exchange membrane fuel cells using a
  model-based method.
\newblock \emph{IFAC-PapersOnLine}, 50\penalty0 (1):\penalty0 4757--4762, 2017.

\bibitem[Bressel et~al.(2016)Bressel, Hilairet, Hissel, and
  Bouamama]{bressel2016remaining}
Mathieu Bressel, Mickael Hilairet, Daniel Hissel, and Belkacem~Ould Bouamama.
\newblock Remaining useful life prediction and uncertainty quantification of
  proton exchange membrane fuel cell under variable load.
\newblock \emph{IEEE Transactions on Industrial Electronics}, 63\penalty0
  (4):\penalty0 2569--2577, 2016.

\bibitem[Jouin et~al.(2014)Jouin, Gouriveau, Hissel, P{\'e}ra, and
  Zerhouni]{jouin2014prognostics}
Marine Jouin, Rafael Gouriveau, Daniel Hissel, Marie-C{\'e}cile P{\'e}ra, and
  Noureddine Zerhouni.
\newblock Prognostics of pem fuel cell in a particle filtering framework.
\newblock \emph{International Journal of Hydrogen Energy}, 39\penalty0
  (1):\penalty0 481--494, 2014.

\bibitem[Napoli et~al.(2013)Napoli, Ferraro, Sergi, Brunaccini, and
  Antonucci]{napoli2013data}
Giuseppe Napoli, Marco Ferraro, Francesco Sergi, Giovanni Brunaccini, and
  Vincenzo Antonucci.
\newblock Data driven models for a pem fuel cell stack performance prediction.
\newblock \emph{International journal of hydrogen energy}, 38\penalty0
  (26):\penalty0 11628--11638, 2013.

\bibitem[Silva et~al.(2014)Silva, Gouriveau, Jemei, Hissel, Boulon, Agbossou,
  and Steiner]{silva2014proton}
RE~Silva, Rafael Gouriveau, Samir Jemei, Daniel Hissel, Lo{\"\i}c Boulon, Kodjo
  Agbossou, and N~Yousfi Steiner.
\newblock Proton exchange membrane fuel cell degradation prediction based on
  adaptive neuro-fuzzy inference systems.
\newblock \emph{International Journal of Hydrogen Energy}, 39\penalty0
  (21):\penalty0 11128--11144, 2014.

\bibitem[Javed et~al.(2016)Javed, Gouriveau, Zerhouni, and
  Hissel]{javed2016prognostics}
Kamran Javed, Rafael Gouriveau, Noureddine Zerhouni, and Daniel Hissel.
\newblock Prognostics of proton exchange membrane fuel cells stack using an
  ensemble of constraints based connectionist networks.
\newblock \emph{Journal of Power Sources}, 324:\penalty0 745--757, 2016.

\bibitem[Liu et~al.(2019)Liu, Li, Chen, Yan, Qiu, and Cao]{liu2019remaining}
Jiawei Liu, Qi~Li, Weirong Chen, Yu~Yan, Yibin Qiu, and Taiqiang Cao.
\newblock Remaining useful life prediction of pemfc based on long short-term
  memory recurrent neural networks.
\newblock \emph{International Journal of Hydrogen Energy}, 44\penalty0
  (11):\penalty0 5470--5480, 2019.

\bibitem[Wang et~al.(2020)Wang, Mamo, and Cheng]{wang2020bi}
Fu-Kwun Wang, Tadele Mamo, and Xiao-Bin Cheng.
\newblock Bi-directional long short-term memory recurrent neural network with
  attention for stack voltage degradation from proton exchange membrane fuel
  cells.
\newblock \emph{Journal of Power Sources}, 461:\penalty0 228170, 2020.

\bibitem[Zuo et~al.(2021)Zuo, Cheng, and Zhang]{zuo2021degradation}
Bin Zuo, Junsheng Cheng, and Zehui Zhang.
\newblock Degradation prediction model for proton exchange membrane fuel cells
  based on long short-term memory neural network and savitzky-golay filter.
\newblock \emph{International Journal of Hydrogen Energy}, 46\penalty0
  (29):\penalty0 15928--15937, 2021.

\bibitem[Long et~al.(2022)Long, Wu, Li, and Li]{long2022novel}
Bing Long, Kunping Wu, Pengcheng Li, and Meng Li.
\newblock A novel remaining useful life prediction method for hydrogen fuel
  cells based on the gated recurrent unit neural network.
\newblock \emph{Applied Sciences}, 12\penalty0 (1):\penalty0 432, 2022.

\bibitem[Vichard et~al.(2020)Vichard, Harel, Ravey, Venet, and
  Hissel]{vichard2020degradation}
Lo{\"\i}c Vichard, Fabien Harel, Alexandre Ravey, Pascal Venet, and Daniel
  Hissel.
\newblock Degradation prediction of pem fuel cell based on artificial
  intelligence.
\newblock \emph{International Journal of Hydrogen Energy}, 45\penalty0
  (29):\penalty0 14953--14963, 2020.

\bibitem[Zhou et~al.(2017)Zhou, Gao, Breaz, Ravey, and
  Miraoui]{zhou2017degradation}
Daming Zhou, Fei Gao, Elena Breaz, Alexandre Ravey, and Abdellatif Miraoui.
\newblock Degradation prediction of pem fuel cell using a moving window based
  hybrid prognostic approach.
\newblock \emph{Energy}, 138:\penalty0 1175--1186, 2017.

\bibitem[Cheng et~al.(2018)Cheng, Zerhouni, and Lu]{cheng2018hybrid}
Yujie Cheng, Noureddine Zerhouni, and Chen Lu.
\newblock A hybrid remaining useful life prognostic method for proton exchange
  membrane fuel cell.
\newblock \emph{International Journal of Hydrogen Energy}, 43\penalty0
  (27):\penalty0 12314--12327, 2018.

\bibitem[Xie et~al.(2020)Xie, Ma, Pu, Xu, Zhao, and Huangfu]{xie2020prognostic}
Renyou Xie, Rui Ma, Sicheng Pu, Liangcai Xu, Dongdong Zhao, and Yigeng Huangfu.
\newblock Prognostic for fuel cell based on particle filter and recurrent
  neural network fusion structure.
\newblock \emph{Energy and AI}, 2:\penalty0 100017, 2020.

\bibitem[Ma et~al.(2021)Ma, Xie, Xu, Huangfu, and Li]{ma2021hybrid}
Rui Ma, Renyou Xie, Liangcai Xu, Yigeng Huangfu, and Yuren Li.
\newblock A hybrid prognostic method for pemfc with aging parameter prediction.
\newblock \emph{IEEE Transactions on Transportation Electrification},
  7\penalty0 (4):\penalty0 2318--2331, 2021.

\bibitem[Xia et~al.(2022)Xia, Wang, Ma, Zhu, Li, Tao, and Tian]{xia2022hybrid}
Zetao Xia, Yining Wang, Longhua Ma, Yang Zhu, Yongjie Li, Jili Tao, and
  Guanzhong Tian.
\newblock A hybrid prognostic method for proton-exchange-membrane fuel cell
  with decomposition forecasting framework based on aekf and lstm.
\newblock \emph{Sensors}, 23\penalty0 (1):\penalty0 166, 2022.

\bibitem[Bhattacharya(2015)]{bhattacharya2015water}
Prashant~K. Bhattacharya.
\newblock Water flooding in the proton exchange membrane fuel cell.
\newblock \emph{Directions}, 15\penalty0 (1), 2015.

\bibitem[Ralli{\`e}res(2011)]{rallieres:tel-00819317}
Olivier Ralli{\`e}res.
\newblock \emph{{Mod{\'e}lisation et caract{\'e}risation de Piles A Combustible
  et Electrolyseurs PEM}}.
\newblock Theses, Toulouse INP, November 2011.
\newblock URL \url{https://theses.hal.science/tel-00819317}.

\bibitem[Mardle et~al.(2021)Mardle, Cerri, Suzuki, and
  El-Kharouf]{mardle2021examination}
Peter Mardle, Isotta Cerri, Toshiyuki Suzuki, and Ahmad El-Kharouf.
\newblock An examination of the catalyst layer contribution to the disparity
  between the nernst potential and open circuit potential in proton exchange
  membrane fuel cells.
\newblock \emph{Catalysts}, 11\penalty0 (8):\penalty0 965, 2021.

\bibitem[Fontès(2005)]{fontes2005}
Guillaume Fontès.
\newblock \emph{Modélisation et caractérisation de la pile PEM pour l'étude
  des interactions avec les convertisseurs statiques}.
\newblock Theses, Toulouse INP, September 2005.
\newblock URL \url{http://www.theses.fr/2005INPT020H}.

\bibitem[El~Aabid(2020)]{aabid2020}
Sami El~Aabid.
\newblock \emph{Méthode basée modèle pour le diagnostic de l'état de santé
  d'une pile à combustible PEMFC en vue de sa maintenance}.
\newblock Theses, Toulouse INP, January 2020.
\newblock URL \url{http://www.theses.fr/2020INPT0011}.

\bibitem[Mansouri et~al.(2024)Mansouri, Alenabi, and
  Gavagsaz-ghoachani]{mansouri2024investigating}
Amirhosein Mansouri, Seyed~Ali Alenabi, and Roghayeh Gavagsaz-ghoachani.
\newblock Investigating performance of hydrogen fuel cells in different charge
  transfer coefficients and its effect on maximum powerpoint.
\newblock \emph{Iranica Journal of Energy \& Environment}, 15\penalty0
  (3):\penalty0 311--318, 2024.

\bibitem[Mor{\'e}(2006)]{more2006levenberg}
Jorge~J Mor{\'e}.
\newblock The levenberg-marquardt algorithm: implementation and theory.
\newblock In \emph{Numerical analysis: proceedings of the biennial Conference
  held at Dundee, June 28--July 1, 1977}, pages 105--116. Springer, 2006.

\bibitem[Kalman(1960)]{kalman1960new}
Rudolph~Emil Kalman.
\newblock A new approach to linear filtering and prediction problems.
\newblock \emph{Journal of Basic Engineering}, 82\penalty0 (1):\penalty0 35--45
  (11 pages), 1960.

\bibitem[Ribeiro(2004)]{ribeiro2004kalman}
Maria~Isabel Ribeiro.
\newblock Kalman and extended kalman filters: Concept, derivation and
  properties.
\newblock \emph{Institute for Systems and Robotics}, 43\penalty0 (46):\penalty0
  3736--3741, 2004.

\bibitem[CJC(2003)]{kruger2003constrained}
Kruger CJC.
\newblock Constrained cubic spline interpolation.
\newblock \emph{Chemical Engineering Applications}, 1\penalty0 (1), 2003.

\bibitem[Robert et~al.(1999)Robert, Casella, and Casella]{robert1999monte}
Christian~P Robert, George Casella, and George Casella.
\newblock \emph{Monte Carlo statistical methods}, volume~2.
\newblock Springer, 1999.

\end{thebibliography}

\end{document}